\documentclass[prodmode]{acmsmall} 

\usepackage[ruled]{algorithm2e}

\SetAlFnt{\small}
\SetAlCapFnt{\small}
\SetAlCapNameFnt{\small}
\SetAlCapHSkip{0pt}
\IncMargin{-\parindent}

\begin{document}

\markboth{M. Kejriwal and D. P. Miranker}{Experience: Type alignment on DBpedia and Freebase}

\title{Experience: Type alignment on DBpedia and Freebase}
\author{MAYANK KEJRIWAL
\affil{University of Texas at Austin}
DANIEL P. MIRANKER
\affil{University of Texas at Austin}
}

\begin{abstract}

Linked Open Data exhibits growth in both volume and variety of published data. Due to this variety, instances of many different types (e.g. Person) can be found in published datasets. \emph{Type alignment} is the problem of automatically matching types (in a possibly many-many fashion) between two such datasets.
Type alignment is an important preprocessing step in \emph{instance matching}. Instance matching concerns identifying pairs of instances referring to the same underlying entity. By performing type alignment \emph{a priori}, only instances conforming to aligned types are processed together, leading to significant savings. This article describes a type alignment experience with two large-scale \emph{cross-domain} RDF knowledge graphs, DBpedia and Freebase, that contain hundreds, or even thousands, of unique types. Specifically, we present a  MapReduce-based type alignment algorithm and show that there are at least three reasonable ways of evaluating type alignment within the larger context of instance matching. We comment on the consistency of those results, and note some general observations for researchers evaluating similar algorithms on cross-domain graphs.
\end{abstract}

%
%


%
%



\acmformat{Mayank Kejriwal and Daniel P. Miranker, 2016. Experience: Type alignment on DBpedia and Freebase.}

\begin{bottomstuff}
This work is supported by the National Science Foundation

Author's addresses: M. Kejriwal {and} D. P. Miranker, Department of Computer Science,
University of Texas at Austin.
\end{bottomstuff}

\maketitle

\section{Introduction}\label{introduction}
\begin{figure}
\centerline{\includegraphics[height=5.3cm, width=12.0cm]{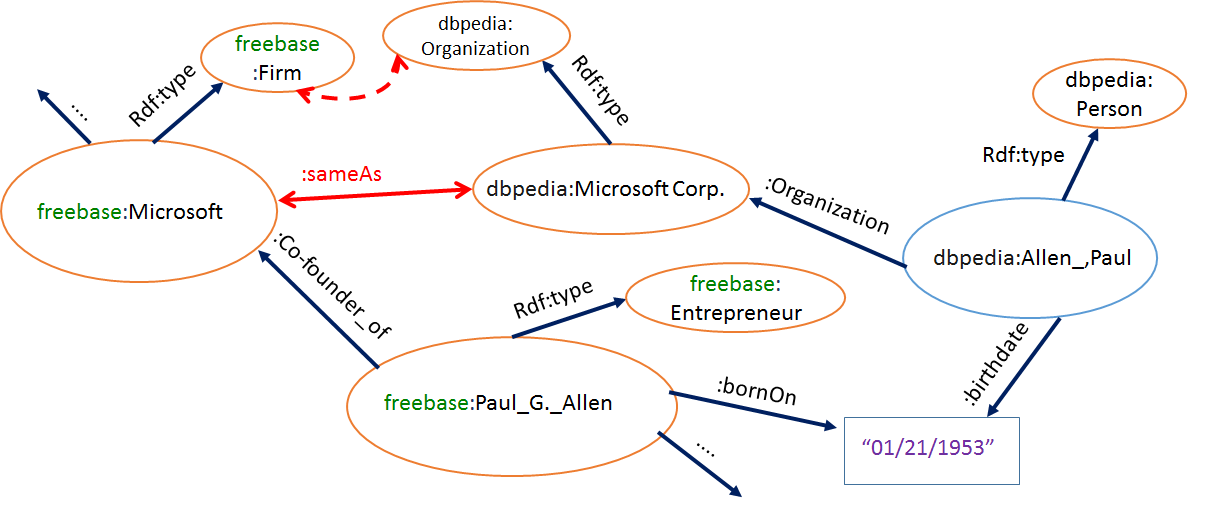}}
\caption{Fragments of two RDF Knowledge Graphs (KGs), Freebase and DBpedia, with an illustration of both the \emph{type alignment} problem (\emph{dotted} edge) and the \emph{instance matching} problem (\emph{:sameAs} edge)}
\label{rdfExample}
\end{figure}

In the last decade, a variety of \emph{knowledge graphs} (KGs) have been published on the Web of Linked Data \cite{linkeddata}. Such graphs typically encode facts in the form of \emph{sets of triples}, most commonly employing the Resource Description Framework (RDF) data model \cite{rdf}. Figure \ref{rdfExample} illustrates examples of two interconnected RDF KGs (Freebase and DBpedia) currently available on the Web. Each triple in the KG \emph{visualized} as a directed, labeled edge. Using RDF, along with the four principles of \emph{Linked Data}, practitioners can publish KGs in a distributed manner across the Web, and access them using standard World Wide Web protocols \cite{linkeddata}. This framework has been widely adopted since its founding, forming a core component of the Semantic Web ecosystem \cite{bernerssemantic}.

The \emph{fourth} Linked Data principle states that datasets should not exist in silos but be linked to each other. Success of the principle has directly led to the growth of Linked Open Data\footnote{\url{http://linkeddata.org/}}, an initiative that started in 2007 and contains many open datasets in domains ranging from publications to social media \cite{linkeddata}. Technically, instances of many different \emph{types} have been published, with an ontology often used to establish a hierarchy between types. Operationally, an \emph{rdf:type} property is used to indicate the type of an instance. In Figure \ref{rdfExample}, the instance \emph{Microsoft} in the Freebase dataset has type \emph{Firm}.

Type alignment is the problem of automatically matching types between two or more datasets. In Figure \ref{rdfExample}, for example, the two types \emph{Firm} and \emph{Organization} in datasets Freebase and DBpedia respectively, are equivalent and should be aligned, as indicated by the dashed bidirectional line. At the same time, there is a clear \emph{sub-type} (but not equivalence) relationship between \emph{freebase: Entrepreneur} and \emph{dbpedia: Person}. Whether these two types should be aligned depends on the application.

One application that directly relies on type alignment is \emph{instance matching}, which concerns identifying entities that refer to the same underlying entity and linking them using \emph{:sameAs} links (Figure \ref{rdfExample}). In turn, this helps practitioners obey the fourth Linked Data principle. Deducing automatic solutions to instance matching continues to be an active area of research in the broader Artificial Intelligence community (Section \ref{relatedwork}).

Due to the growth of \emph{cross-domain} datasets such as DBpedia\footnote{\url{http://wiki.dbpedia.org/}} and Freebase\footnote{\url{http://www.freebase.com/}}, the problem has manifested itself as a core Big Data problem \cite{kejriwalENS}. Such datasets are mostly derived from encyclopedias (e.g. Wikipedia), contain hundreds (DBpedia) or even thousands (Freebase) of different types \cite{KG}, and have invited multiple applications \cite{okkam}, \cite{kgi}.

The heterogeneity of these graphs suggest that type alignment should either precede a viable instance matching procedure \cite{entitytype}, or that instance matching algorithms should be \emph{schema-free} \cite{jws}. \emph{Scalability} concerns suggest the former option is preferrable \cite{ibm}. In Figure \ref{rdfExample}, aligning \emph{Firm} and \emph{Organization} would preclude wasted comparisons of entities that have incompatible types (e.g. comparing a firm to a person), leading to significant savings\footnote{The figure also shows that, for good \emph{coverage}, a type alignment solution used in an instance matching application must also align types that have sub-type and super-type relationships.}.

Given a type alignment solution and a benchmark test suite, it is straightforward, in principle, to evaluate the correctness of the type alignment solution. This article shows that this assumption is problematic for large-scale cross-domain KGs that contain significant noise. In particular, we detail three possible ways of evaluating a type alignment solution, the end goal being scalable instance matching (Section \ref{typealignment}). The three different ground-truths can yield potentially conflicting results, raising important issues about type alignment \emph{assessment} (Section \ref{experiments}). We use our experience matching instances between DBpedia and Freebase to shed light on these issues, and offer lessons for others attempting a similar task.  

\section{Related Work}\label{relatedwork}
In this article, type alignment is studied within the larger context of instance matching. Thus, we first cover general related work on instance matching, followed by type alignment.
\subsection{Instance Matching}\label{im}
Instance matching has emerged as an important research problem in both the structured and semistructured data communities \cite{recordlinkagesurvey}, \cite{rahmsurvey}. A scalable solution typically involves a \emph{two-step} formulation \cite{datamatching} on \emph{single-type} datasets. The first step, \emph{blocking}, clusters entities into (possibly overlapping) \emph{blocks}, such that entities within blocks are approximately similar to each other and merit further expensive comparisons. In the second step, often called \emph{similarity} or \emph{classification}, only entities sharing blocks are paired, with each pair classified as duplicate or non-duplicate.

Instance matching has been comprehensively surveyed by multiple authors, including Elmagarmid et al. \cite{recordlinkagesurvey} and K{\"o}pcke and Rahm \cite{rahmsurvey}. The problem continues to be an evolving research area in the Semantic Web. Proposed applications include knowledge graph identification \cite{kgi} and semantic search \cite{okkam}. In particular, recent years have seen the increased use of  \emph{minimally supervised} machine learning techniques \cite{eswc}, \cite{machine1}.

\subsection{Type Alignment}\label{ta}
Recently, instance matching practitioners have recognized that datasets often contain instances of \emph{multiple types}, and have employed some form of \emph{ontology matching} to match types prior to instance matching itself \cite{ontologysurvey}. In the \emph{Raven} system, for example, Ngomo et al. generate type and property alignments by framing the problem as an application of \emph{stable marriage} \cite{raven}. We presented a second algorithm that works along similar lines but uses a different, lower-complexity, matching algorithm instead \cite{kejriwaltwo}. A second approach to addressing multiple types, which has been gaining traction recently, is to devise \emph{schema-free} algorithms that do not need to match types before performing instance matching \cite{machine1}, \cite{eswcSN}.

Common among all the different approaches is that there is a single type alignment ground-truth. Thus, given that a system outputs that type A matches type B, the output is either marked as correct (if present in the ground-truth) or incorrect. We show in Section \ref{evalmethod} that this assumption of a single fixed ground-truth may not always be realistic when type alignment is performed within the larger context of instance matching. To the best of our knowledge, this issue, which directly affects quality assessment of a type alignment solution, has not been addressed previously.

Finally, the experience described herein concerns instance matching designed for cross-domain knowledge graphs (KGs) such as DBpedia and Freebase. A comparative survey of such KGs was recently provided by F{\"a}rber et al [2015]. To the best of our knowledge, only the work by \cite{ibm} has handled type alignment tasks on such graphs in an unsupervised distributed setting.

\section{Type Alignment: A MapReduce-based algorithm}\label{typealignment}

\begin{figure}
\centerline{\includegraphics[height=6.0cm, width=13.0cm]{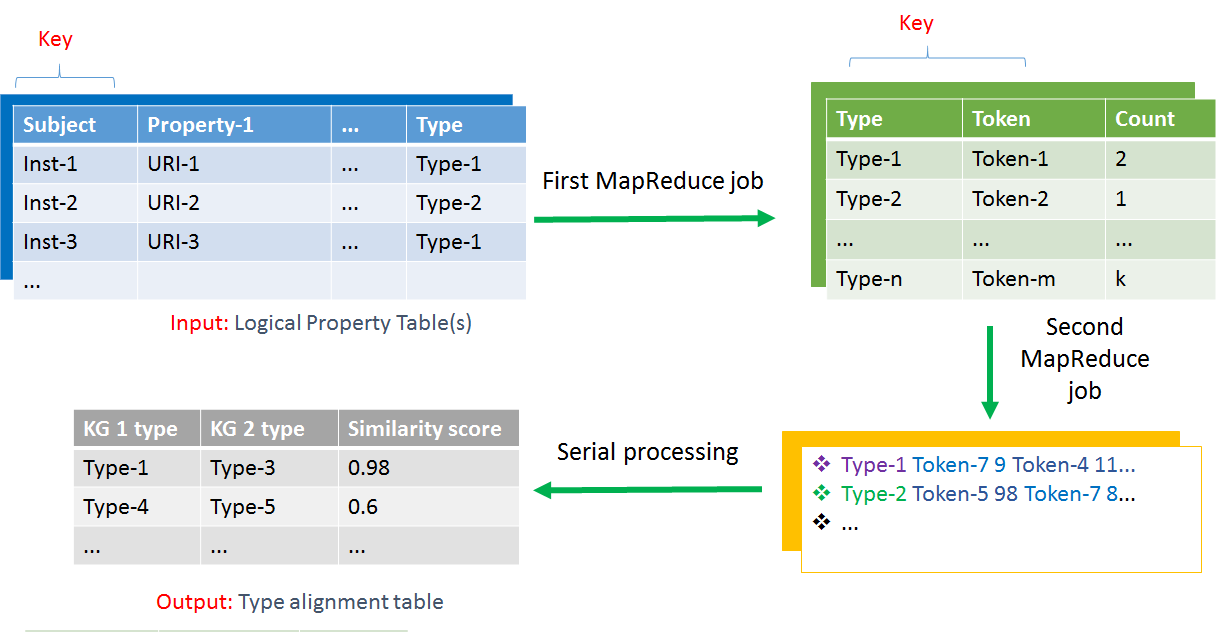}}
\caption{The token-based type alignment algorithm. The output is a table that lists type alignments between two knowledge graphs (KGs) with corresponding scores according to a well-defined similarity measure (Jaccard, Generalized Jaccard or Log-Term Frequency)}
\label{algo1}
\end{figure}

While many type alignment systems exist (typically as sub-algorithms in ontology matching architectures \cite{ontologysurvey}), a key need only recently addressed is \emph{scalability}. Since our instance matcher is designed for large-scale KGs stored in a distributed file system as use-cases, and we used the public cloud  for experiments (Section \ref{experiments}), only those type aligners were considered that could be scaled and implemented in MapReduce \cite{mapreduce}. A second consideration was that the type aligner had to be \emph{instance-driven} rather than \emph{structural}, owing to the shallow, \emph{schema-free} nature of cross-domain ontologies on Linked Open Data \cite{linkeddata}.

Pilot experiments on a development set showed a \emph{token-based} approach to yield both fast and robust results. The algorithm is illustrated in Figure \ref{algo1}. The algorithm assumes that each instance in each KG is encoded in a NoSQL format (such as JSON) and contains the same information set as a row in a \emph{logical} property table encoding \cite{kejriwaltwo}, shown in Figure \ref{algo1}. First, a MapReduce job is run separately for each KG (say $A$ and $B$), where a mapper takes an instance $I_A$ as input, converts it into a \emph{set} of tokens using some standard delimiters (e.g. punctuations) and emits a key-value pair of the form \emph{(Type($I_A$)-$token_i$, one)}, where $one$ is a placeholder, \emph{Type($I_A$)} is the type of instance $I_A$ and $token_i$ is an arbitrary token from the set of tokens. Note that if a token occurs multiple times within an instance, it is still counted only once, since we are interested in collecting type-token statistics. The reducer simply counts all type-tokens and emits an intermediate output visualized in Figure \ref{algo1}.

In the second MapReduce job, all the type-tokens, with their counts, are consolidated in a single line\footnote{Just like with the first MapReduce job, the second job is also run separately for each KG.}. Data skew was found to be a significant problem in the original run of the second MapReduce job. To avoid skew, the mapper in the second MapReduce job did not emit tokens that had count less than 5; in the reducer, only the first 30,000 unique tokens emitted by a mapper (per type) were written out to the final consolidated file. 

An advantage of the two-step MapReduce algorithm is that the output is quite compact. Freebase, which is just slightly under 400 GB in uncompressed form, produced a final type-token output of less than 150 MB (less than 10 MB for DBpedia). Due to the compactness, types between two KGs could be matched \emph{serially}, using the tokens and their statistics. Three standard set-based similarity measures (normalized in the [0,1] range) were used: Jaccard, Generalized Jaccard (g-Jaccard), and Log-Term Frequency (or log-TF\footnote{\emph{Inverse} frequencies were not employed, with a rationale provided in Section \ref{experiments}.}). Details for these measures may be found in any standard text on data matching (e.g. \cite{datamatching}); herein, only brief details are provided. Consider two arbitrary types $t_1$ and $t_2$ (from two KGs A and B resp.), each represented by a set of tokens (with counts)\footnote{We use as shorthand the pattern $t_{1|2}=f_{m|n}$ to indicate the  equations $t_1=f_m$ and $t_2=f_n$}: $t_{1|2}:\{<token_1,count_1>,..,<token_{m|n},count_{m|n}>\}$. The Jaccard measure ignores counts completely and is given by the formula $|Tokens_1 \cap Tokens_2|/|Tokens_1 \cup Tokens_2|$, where $Tokens_{1|2}$ represents the set of tokens of $t_{1|2}$, obtained by \emph{neglecting} counts. The other two similarity measures, g-Jaccard and log-TF, first \emph{normalize}\footnote{Note that g-Jaccard and log-TF respectively employ norm-1 and norm-2 normalization.} each count to a value in the range [0.0,1.0].  Denoting the combined set of tokens $Tokens_1 \cup Tokens_2$ by $Tokens$, the g-Jaccard score is computed using the formula $\Sigma_{Tokens} min(count_i^1, count_i^2)/\Sigma_{Tokens} max(count_i^1, count_i^2)$, where $count_i^1$ and $count_i^2$ are the respective counts in types $t_1$ and $t_2$ of some token $token_i \in Tokens$\footnote{If $token_i$ is present in only one type, it is interpreted as having count zero in the missing type.}. Log-tf is the logarithm of the dot product of both count vectors.

Intuitively, the higher the similarity score according to a given measure, the more likely it is that the corresponding types are aligned. Using a threshold, which can be varied to trade-off various metrics against each other (Section \ref{experiments}), a curve can be plotted to visualize empirical type alignment performance against a ground-truth.

\section{Evaluation methodologies (Ground-truths)}\label{evalmethod}
Earlier, we suggested that evaluating type alignment is not a straightforward issue in the larger context of instance matching. We now describe three separate ground-truths that can be constructed to evaluate type alignment in this context, and the rationale behind each construction.

\paragraph{Ground-truth 1} One way to measure the correctness of the type alignment algorithm is by measuring its \emph{utility} for the instance matching task in terms of scalability. That is, the only purpose of type alignment is to make instance matching (between two KGs $A$ and $B$) more scalable without losing recall. For any ground-truth to reflect this, the following assumption must hold: \emph{the type-pair $(T_A,T_B)$ is included in the ground-truth iff there exists an instance pair $(I_A, I_B)$ such that $I_A$ has type $T_A$ and $I_B$ has type $T_B$.} Using this assumption, a ground-truth can be easily constructed by including each pair of types that covers at least one matching instance pair. A version of this ground-truth was used in the work by \cite{ibm}.

\paragraph{Ground-truth 2} While ground-truth 1 makes a reasonable assumption, it ignores the issue of \emph{data skew}. Consider, for example, the type pair \emph{(Person, Footballer)}. While this type pair covers many instance pairs, it hardly serves any practical purpose (from a scalability standpoint), since \emph{Person} covers many sub-types, including non-footballers. A `better' alignment would be \emph{(Football-player, Footballer)}, assuming that \emph{Football-player} is a type in KG $A$ and \emph{Footballer} in KG $B$. Intuitively, the alignment is better because it offers roughly the same coverage (of matching instance pairs) but with greater savings. A proper evaluation of type alignment from this standpoint would utilize the instance matching ground-truth \emph{directly}, bypassing the need for a separate type-alignment ground-truth. Technically, the type alignment would be evaluated in exactly the same way as \emph{blocking}\footnote{This is not intended to \emph{replace} blocking. Once the types are aligned, blocking and similarity (the two-step formulation) are still performed on the instance space covered by each aligned type pair.} in a two-step instance matching formulation (see Section \ref{relatedwork}). The standard blocking metrics (efficiency and coverage) can then be used to evaluate the type alignment. Details are provided in Section \ref{experiments}. This ground-truth represents the `blocking within blocking' paradigm adopted by \cite{typimatch}

\paragraph{Ground-truth 3} Finally, we can also evaluate type alignment as a part of ontology matching, thereby neglecting the context of instance matching. A user would manually create a type alignment set by considering the structural and descriptive properties of a candidate type pair. The rationale is that type alignment is fundamentally different from instance matching, and the ground-truth should reflect this. While ground-truths 1 and 2 are \emph{utilitarian}, this ground-truth is more fundamental. Systems like Raven typically rely on this ground-truth \cite{raven}.   

Note that each of the three ground-truths is evaluated on a different set of metrics, some of which may not be directly comparable. More details are provided in Section \ref{results}.

\section{Evaluations}\label{experiments}

\subsection{Setup}
Evaluations were performed on DBpedia and Freebase, on Microsoft Azure cloud infrastructure\footnote{\url{https://azure.microsoft.com/en-us/}}. The overall goal of our system was to perform minimally supervised instance matching between two input graphs. Within that larger context, the scope of this article is limited to evaluating, on DBpedia and Freebase, the type alignment algorithm (presented in Section \ref{typealignment}) using the three ground-truths described earlier in Section \ref{evalmethod}. A serial version of the overall system, which did not include type alignment as a separate component, has already been published\footnote{A cloud-based version including type alignment is currently under review.} \cite{jws}.

The evaluation was setup as follows. First, the publicly available N-triples files containing DBpedia and Freebase facts were downloaded\footnote{We downloaded the versions available in early August. Freebase has not been updated since then, but DBpedia continues to be frequently updated.} and stored in Microsoft Azure cloud storage as blobs. For DBpedia, two separate triples files had to be downloaded and merged into a single file. The first of these described instance type information\footnote{We only considered \url{dbpedia.org/ontology} types.}, while the second described instance properties (facts). There were 3.279 million unique subjects, 67.1 million unique tripes, and 417 unique types in the merged file.  For Freebase, only one triples file was available and contained 121.629 million unique subjects, 3.023 billion unique triples and 4811 unique types. We also downloaded a third-party file describing approximately 3.3 million \emph{:sameAs} links between the instances.

We ran a serialization algorithm in MapReduce to convert the triples files to the NoSQL format that is required by the MapReduce steps of the type alignment algorithm. For all MapReduce algorithms, we used between 6 and 15 quad-core data nodes\footnote{Technically, A3 HDInsight nodes: \url{https://azure.microsoft.com/en-us/pricing/details/hdinsight/}} with 7 GB of memory  and 285 GB disk size each. 
The total MapReduce serialization time for DBpedia was 14 minutes, and for Freebase, around 4.5 hours.
Once the data was formatted correctly, the type alignment algorithms were run for both DBpedia and Freebase, the respective run-times (for the chained MapReduce job) being 13 minutes and 6 hours. The consolidated type-tokens files (see Figure \ref{algo1}) were downloaded to a local machine. Further processing, including analyses with respect to the different ground-truths described in Section \ref{evalmethod}, was done serially in near-instantanous time. These serial evaluations are described in the next section.

\subsection{Results and Discussion}\label{results}
\begin{figure}
\centerline{\includegraphics[height=4.8cm, width=13.0cm]{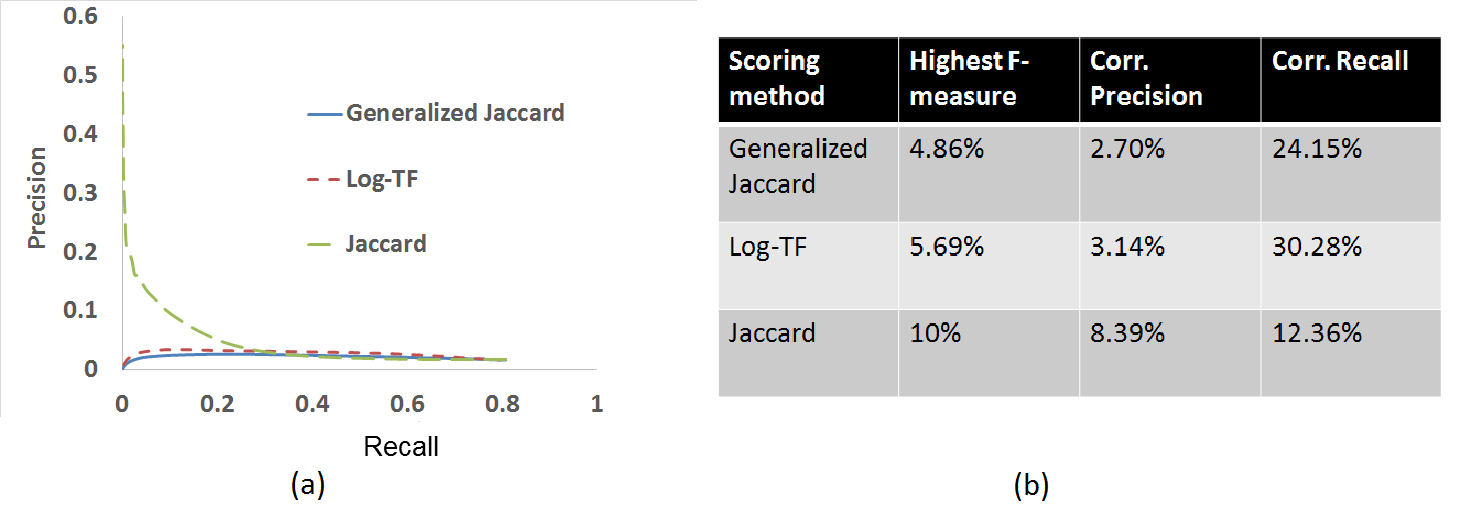}}
\caption{Results using ground-truth 1. (a) measures precision vs. recall by varying a threshold, while in (b), the highest-obtained F-measures are recorded, along with the corresponding values of precision and recall at which the F-measure was obtained}
\label{evalFig1}
\end{figure}

\begin{figure}
\centerline{\includegraphics[height=4.8cm, width=13.5cm]{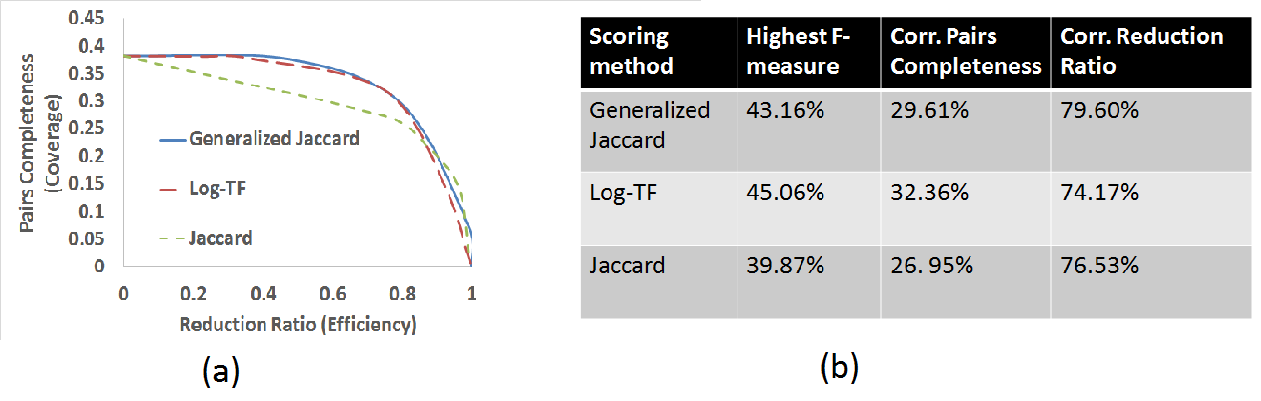}}
\caption{Results using ground-truth 2. (a) measures Pairs Completeness vs. Reduction Ratio by varying a threshold, while in (b), the highest-obtained F-measures are recorded, along with the corresponding values of Pairs Completeness and Reduction Ratio at which the F-measure was obtained}
\label{evalFig2}
\end{figure}

\begin{figure}
\centerline{\includegraphics[height=4cm, width=13.0cm]{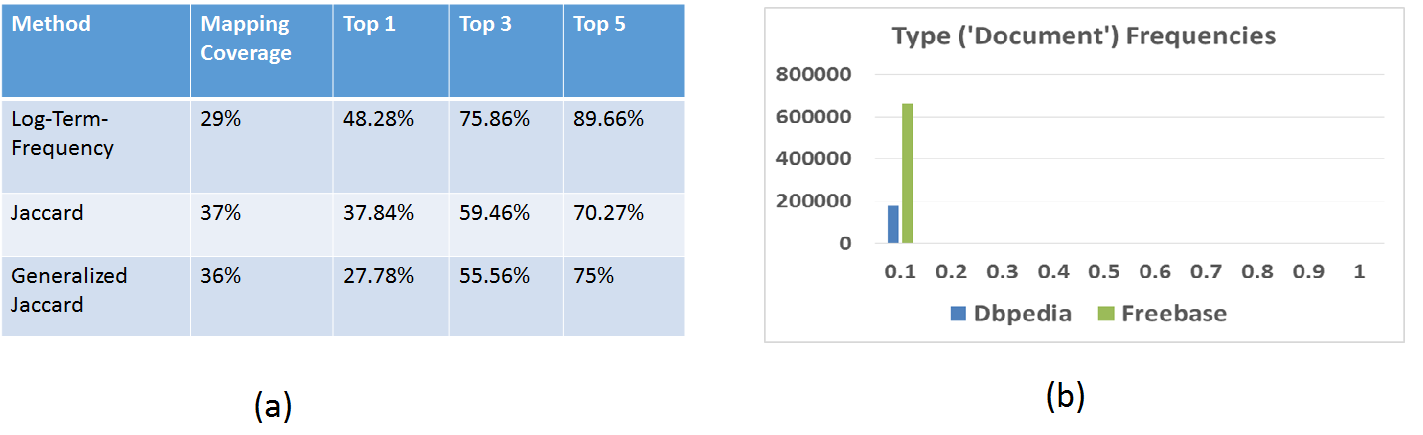}}
\caption{Results using ground-truth 3. (a) illustrates the coverage results, while (b) provides an explanation for why inverse frequencies are not necessary for this problem}
\label{evalFig3}
\end{figure}

Figures \ref{evalFig1}, \ref{evalFig2} and \ref{evalFig3} respectively illustrate the results for ground-truths 1, 2 and 3.
Figure \ref{evalFig1} (a) illustrates the results (on ground-truth 1) for the three similarity measures earlier described in Section \ref{typealignment}. Two metrics, \emph{precision} and \emph{recall}, were being measured in this experiment. Precision measures the ratio of true positives to the sum of true positives and false \emph{positives}, whereas recall measures the ratio of true positives to the sum of true positives and false \emph{negatives}. We obtained the graph by varying the threshold at which the type pairs were declared to be aligned. The results in \ref{evalFig1} (a) show that, in general, the Jaccard similarity measure outperforms the other measures, although overall performance on all three are quite low. The highest F-measure\footnote{Defined as the harmonic mean of precision and recall.} obtained in the experiment was also low, but with corresponding recall often an order of magnitude higher than corresponding precision. Despite this, the absolute recall is around 30\% or less, which indicates that either the ground-truth or the alignment solution should be a cause for concern. We analyzed the ground-truth and found that there were many type pairs that covered instance pairs, but were \emph{semantically} misleading. For example, \emph{Football-player} was found to match the type \emph{Criminal}, since there are football players who are convicted criminals. Thus, ground-truth 1 was sensitive to the \emph{:sameAs} links in the dataset, making its utility on noisy cross-domain KGs, questionable. 

As noted in Section \ref{evalmethod}, ground-truth 2 is `virtual', in that type alignment is evaluated indirectly by treating it akin to \emph{blocking}, and computing two blocking metrics, Pairs Completeness (PC) and Reduction Ratio (RR), for a given type alignment solution. Given an instance matching ground-truth, which is fixed, both metrics can be computed as follows. Let $I_{all}$ denote the set of all instance pairs covered by the retrieved type alignments, $I_{G}$ denote the \emph{duplicates} ground-truth and $I_{TP} = I_{G} \cap I_{all}$ denote the set of retrieved true positives (duplicate instance pairs). Intuitively, we may think of each type alignment as being similar to a block. From that perspective, PC measures the recall of a type alignment solution on instance matching and is the ratio $|I_{TP}|/|I_{G}|$. RR, a  measure of blocking efficiency, is computed as $1.0-|I_{G}|/|\Omega|$, where $\Omega=I_A \times I_B$; that is the \emph{exhaustive} space of instance pairs in knowledge graphs $A$ and $B$. Intuitively, the higher the RR\footnote{In practice, we often compute a lower bound for RR, assuming the blocks (or type alignments) are \emph{overlapping}, that is, cover an instance pair multiple times. For such cases, it is possible to improve RR even further by deduplicating the candidate set of instance pairs prior to instance matching. In the case of DBpedia and Freebase, where the exhaustive space is of the order of hundreds of trillions, such deduplication is infeasible.} the more we have managed to cut down on overall instance matching complexity through type alignment.

Figure \ref{evalFig2} illustrates the results for ground-truth 2. (a) shows that, unlike with ground-truth 1, the Jaccard measure is not as effective as the other two. (b) shows that Log-TF achieves the highest F-measure at 45.06\% with corresponding PC at 32.36\%,  and corresponding RR at 74.17\%. Although the interpretations of the results obtained in the previous experiment and this experiment are separate, it is worth noting that the numbers (with respect to ground-truth 2) are a lot better. This lends credence to the previous observations made about both data skew (in Section \ref{evalmethod}), and noise in type alignment ground-truths (in the early part of this section). A worrying commonality between both experiments is the low performance (about 20-40\%) on recall-centric measures (PC in this experiment).  We comment more on this issue below.

Ground-truth 3 was formed by using manual expertise to determine whether two types were aligned. Such expertise used specific cues, such as the names of the types, as well as accompanying descriptions. Since the space of all possible type alignments is well over 200,000 for Freebase and DBpedia, we followed an information retrieval approach to construct this ground-truth. We randomly sampled 100 Freebase types, and obtained the top 10 DBpedia types per Freebase type by using the scores output by our type alignment solution (for all three similarity measures).  We manually went through the list of 100 Freebase types (and the corresponding list of top 10 DBpedia retrievals per type) and marked the ones that we deemed as aligned. Four calculations were subsequently performed. First, we computed the \emph{mapping coverage} by calculating the ratio of Freebase types that had an aligned DBpedia type in the top 10. Next, for the covered types, we computed the top 1, top 3 and top 5 coverage, where the top k coverage is defined as follows. We count the number of Freebase types that had an aligned DBpedia type\footnote{Multiple DBpedia alignments per Freebase type were only counted once.} in the top k. This count was divided by the mapping coverage, as earlier obtained. The results are recorded in Figure \ref{evalFig3} (a).  

The results show a fair degree of consistency between the similarity measures. While Jaccard performs well on overall mapping coverage, Log-TF has better top-k coverage for low k. We note again a commonality between all three experimental runs: recall-centric measures (mapping coverage in this experiment) tend to stay in the 20-40\% range. This severely limits the final recall that the overall instance matching system can hope to achieve.

We end this section with a note on why including \emph{global} components (such as an \emph{inverse} type-token frequency) in the \emph{local} similarity measures described thus far did not prove to be empirically helpful. For example, Log-TF can be easily supplemented with an inverse frequency term that \emph{penalizes} tokens present in \emph{many} types. Figure \ref{evalFig3} (b) shows that, in fact, the vast majority of frequencies are in the [0.0,0.1] range, meaning that tokens are highly unique to their types. We believe that this is a consequence of the measures we took in our type alignment algorithm to eliminate data skew. This particular finding is important because considerable memory resources are saved if the similarity measures are local. In particular, a full pass over the data is not required, as it would have been if a token dictionary spanning the entire dataset had to be compiled.
\section{Lessons and Future Work}\label{conclusion}
One commonality between the three methodologies was that the metrics computing recall or coverage tended to be consistent (even though they have different interpretations) across all evaluations. It was also quite low, typically below 40\%. We noted in the early part of Section \ref{experiments} that this can only happen due to noise in the dataset, or the inadequacy of the type alignment solution. We ran controlled experiments with our solution on other datasets, both real and synthetic, and achieved excellent performance. This sign was encouraging, but still leaves open the possibility that our solution may have been inadequate for the specific use-case. To settle this question, we did a post-mortem error analysis, where for all three methodologies, we located examples of cases that were `missed' by the alignment solution and led to loss in coverage. We found significant amounts of noise to be present (an example was provided earlier concerning football-players and criminals); in many cases `compatible' types had nothing in common at all, including discriminative tokens or name-based simliarities. This provides some evidence that the schema-free nature of Linked Open Data makes the linking of LOD datasets challenging, not just from a structural perspective, but also from a quality perspective. 

Perhaps a more sobering lesson derived from this experience is that, in many use-cases, `assuming away' type alignment as a solved problem (prior to instance matching) is not viable in the real-world. This finding has precedence in the Relational Database literature, where many record linkage practitioners often assumed away the schema matching problem, despite its documented difficulties \cite{recordlinkagesurvey}. In addition, this article also sought to highlight that assuming away a type alignment \emph{ground-truth} is also problematic. While some of the results were optimistic (e.g. Figure \ref{evalFig3} (a)), others were pessimistic (Figure \ref{evalFig1}). Clearly, the way that the alignment system is evaluated affects the broad conclusions that can be drawn about system performance.

Currently, we are looking to supplement type alignment with more robust schema-free approaches. Such approaches would treat the type alignments as fuzzy signals, and use them as features in the instance matching process rather than hard constraints. Early results on this approach have been promising. Also in progress are additional experiments on a third cross-domain graph, \emph{Yago}, using both DBpedia and Freebase as the second dataset.

\bibliographystyle{ACM-Reference-Format-Journals}
\bibliography{acmsmall-sample-bibfile}
                            

\end{document}